\documentclass[epj]{webofc}
\usepackage[varg]{txfonts}   
\usepackage{lineno}

\newcommand{\pp}{\ensuremath{pp}}
\newcommand{\pt}{\ensuremath{p_{\mathrm{T}}}}

\woctitle{International Conference on New Frontiers in Physics 2013}

\begin{document}

\title{Pion femtoscopy measurements in ALICE at the LHC}

\author{$\L$ukasz Kamil Graczykowski\inst{1}\fnsep\thanks{\email{lgraczyk@if.pw.edu.pl}} (for the ALICE Collaboration)}

\institute{Faculty of Physics, Warsaw University of Technology, Koszykowa 75, 00-662 Warsaw, Poland}

\abstract{
We present the results of two-pion Bose-Einstein correlations measured in Pb--Pb collisions at a center-of-mass energy $\sqrt{s_{\mathrm{NN}}}=2.76$~TeV recorded by ALICE at the Large Hadron Collider. These types of correlations allow to extract, using the technique of femtoscopy (also known as Hanburry-Brown Twiss interferometry, or shortly HBT), the space-time characteristics of the source from the correlation calculated as a function of the pair momentum difference. The femtoscopic analysis was performed using both the Spherical Harmonics decomposition and the standard 3D Cartesian representation of the correlation function. The source sizes in three dimensions, the HBT radii, were extracted by fitting the experimental correlation functions. The resulting dependencies of the radii as a function of centrality and pair transverse momentum are shown. The results indicate the existence of a flowing medium and provide constraints on existing dynamical models. The ALICE Pb--Pb HBT radii are also compared to the \pp~analysis and other heavy-ion experiments in order to test the multiplicity scaling between different systems.
}

\maketitle
\section{Introduction}
\label{intro}
The Large Hadron Collider (LHC) has started operating in 2009 and has delivered \pp~collisions at center-of-mass energies of $\sqrt{s}=0.9$~TeV, $\sqrt{s}=2.76$~TeV, $\sqrt{s}=7$~TeV and $\sqrt{s}=8$~TeV as well as Pb--Pb collisions at a center-of-mass energy of $\sqrt{s_{\mathrm{NN}}}=2.76$~TeV and p--Pb collisions at $\sqrt{s_{\mathrm{NN}}}=5.02$~TeV. The main goal of A Large Ion Collider Experiment (ALICE) \cite{Aamodt:2008zz}, one of the experiments at the LHC, is the study of the Quark-Gluon Plasma (QGP) which is postulated to be produced in the high energy collisions of heavy-ions. One of the tools used to analyze the properties of QGP is the technique of femtoscopy \cite{Kopylov:1972qw,Kopylov:1974uc}.

Pion femtoscopy (also referred to as Hanburry-Brown Twiss interferometry, or shortly HBT) employs Bose-Einstein quantum statistics in order to measure the sizes of the particle emitting region. These sizes, called the HBT radii, are directly proportional to the width of the enhancement of two-pion femtoscopic correlation function at low relative momentum. Two important features have been revealed by the studies performed for heavy-ion collisions at lower energies \cite{Adams:2005dq,Adcox:2004mh,Back:2004je,Arsene:2004fa}: firstly, the 3-dimensional HBT radii in the Longitudinally Co-Moving System (LCMS)~\cite{Danielewicz:2006hi} scale linearly with the cube root of charged particle pseudorapidity density $\left < \mathrm{d}N_{\mathrm{ch}}/\mathrm{d}\eta \right > ^{1/3}$; and secondly, the radii decrease with increasing pair transverse momentum. Such behavior is described by the "homogeneity lengths mechanism" and suggest the presence of the strong collective radial flow in the medium~\cite{Akkelin:1995gh}. Both of these phenomena are expected to be present at the LHC as well.

In this paper we report on the results of two-pion femtoscopic analysis in Pb--Pb collisions at a center-of-mass energy of $\sqrt{s_{\mathrm{NN}}}=2.76$~TeV per nucleon pair recorded by ALICE in November and December of 2010. The Pb--Pb pion femtoscopy results from ALICE limited to 0--5\% centrality were already published in~\cite{Aamodt:2011mr}. We also note that the results of the pion femtoscopic analysis in \pp~collisions at center of mass energies $\sqrt{s}=0.9\ \rm{TeV}$, $\sqrt{s}=2.76\ \rm{TeV}$ and $\sqrt{s}=7\ \rm{TeV}$ in ALICE were published in~\cite{Aamodt:2010jj,Aamodt:2011kd,Graczykowski:2012xz}. Finally, we show the comparison of ALICE minimum bias \pp~results with the Pb--Pb radii as well as with different havy-ion systems from other experiments at lower energies.


\section{Data sample}
\label{sec:datasample}
ALICE has recorded approximately 12 million minimum bias Pb--Pb collisions at $\sqrt{s_{\mathrm{NN}}}=2.76$~TeV. Two ALICE subsystems were used for this analysis: the Time Projection Chamber (TPC) for particle identification and trajectory reconstruction and the VZERO detectors for triggering and centrality estimation. Seven centrality classes which correspond to 0--5\%, 5--10\%, 10--20\%, 20--30\%, 30--40\%, 40--50\%, and 50--60\% of the total inelastic cross section were studied. The details of centrality determination can be found in~\cite{Aamodt:2010cz}.

In this analysis the location of the primary vertex of each event was required to be within $8$~cm of the TPC center. The studies were performed on primary tracks which were selected according to the minimum distance of a track to the primary vertex (so-called Distance of Closest Approach or DCA). Tracks were required to have DCA not greater than $0.2\ \rm{cm}$ in the transverse plane and $0.15\ \rm{cm}$ in the longitudinal direction. They were also required to have the pseudorapidity within the range $|\eta|<0.8$ which corresponds to the uniform TPC acceptance. Only those particles with transverse momentum \pt~range between 0.14 and 2.0~GeV/$c$ were accepted. The selection of pions was performed according to the specific energy loss, $\mathrm{d}E/\mathrm{d}x$, in the TPC. In addition, the track reconstruction algorithm required at least 80 clusters in the TPC for each track (out of maximum of 159 clusters) and the $\chi^2$ per degree of freedom of the Kalman Filter-based fit to be at most 2.

When forming particle pairs, track merging (two tracks reconstructed as one) and splitting (one track reconstructed as two) effects were reduced by applying specific selection criteria (for details see~\cite{Aamodt:2011mr}). The pairs were grouped in 7 ranges of pair transverse momentum $k_\mathrm{T}=|\mathbf{p_{\mathrm{T,1}}}+\mathbf{p_{\mathrm{T,2}}}|/2$: (0.2--0.3), (0.3--0.4), (0.4--0.5), (0.5--0.6), (0.6--0.7), (0.7--0.8) and (0.8--1.0)~GeV/$c$.

\section{Correlation function analysis}
\label{sec:correlationfunctionanalysis}
The femtoscopic correlation function was constructed as a ratio of the signal to the background distributions: 
\begin{equation}
\label{eq:CorrelationFuntion}
C(\mathbf{q})=\frac{N_{pairs}^{mixed}}{N_{pairs}^{signal}} \frac{S(\mathbf{q})}{B(\mathbf{q})},
\end{equation}
where $\mathbf{q}$ is the relative two-pion momentum in LCMS. The signal distribution was constructed from particle pairs coming from the same event:
\begin{equation}
\label{eq:Nsignal}
S(\mathbf{q})=\frac{\mathrm{d}N_{pairs}^{signal}}{\mathrm{d}\mathbf{q}},
\end{equation}
where $N_{pairs}^{signal}$ is the number of pion pairs. 
The background distribution was constructed using the procedure of event mixing, where each particle in the pair comes from a different event and can be expressed as:
\begin{equation}
\label{eq:Nmixed}
B(\mathbf{q})=\frac{\mathrm{d}N_{pairs}^{mixed}}{\mathrm{d}\mathbf{q}},
\end{equation}
where $N_{pairs}^{mixed}$ is the number of pion pairs in the background distribution $B$. In order to improve the background estimation, each event was mixed with ten different events similar in terms of multiplicity and primary vertex location. 

The resulting three-dimensional correlation functions were analyzed both in 3D Cartesian and Spherical Harmonics (SH) representations. The first one is usually projected into $out$, $side$ and $long$ directions of the Longitudinal Co-Moving System (where $long$ is the direction along the beam, $out$ along the pair transverse momentum, and $side$ perpendicular to the other two). The latter technique allows to represent the three-dimensional object as an infinite set of one-dimensional spherical harmonics. The symmetries of the pair distribution make most of the components vanish and the first three of the non-zero ones, $C_0^0$, $C_2^0$ and $C_2^2$, describe the most important information about the correlation \cite{Chajecki:2008vg,Kisiel:2009iw}. The first one, $C_0^0$, is the angle-averaged component reflecting the overall size of the source; the second one,  $C_2^0$, reflects the difference between \emph{transverse} and \emph{long} while the third one, $C_2^2$, measures the difference between \emph{out} and \emph{side}.

\begin{figure}[!ht]
\centering
\begin{minipage}{.45\textwidth}
\includegraphics*[width=\textwidth]{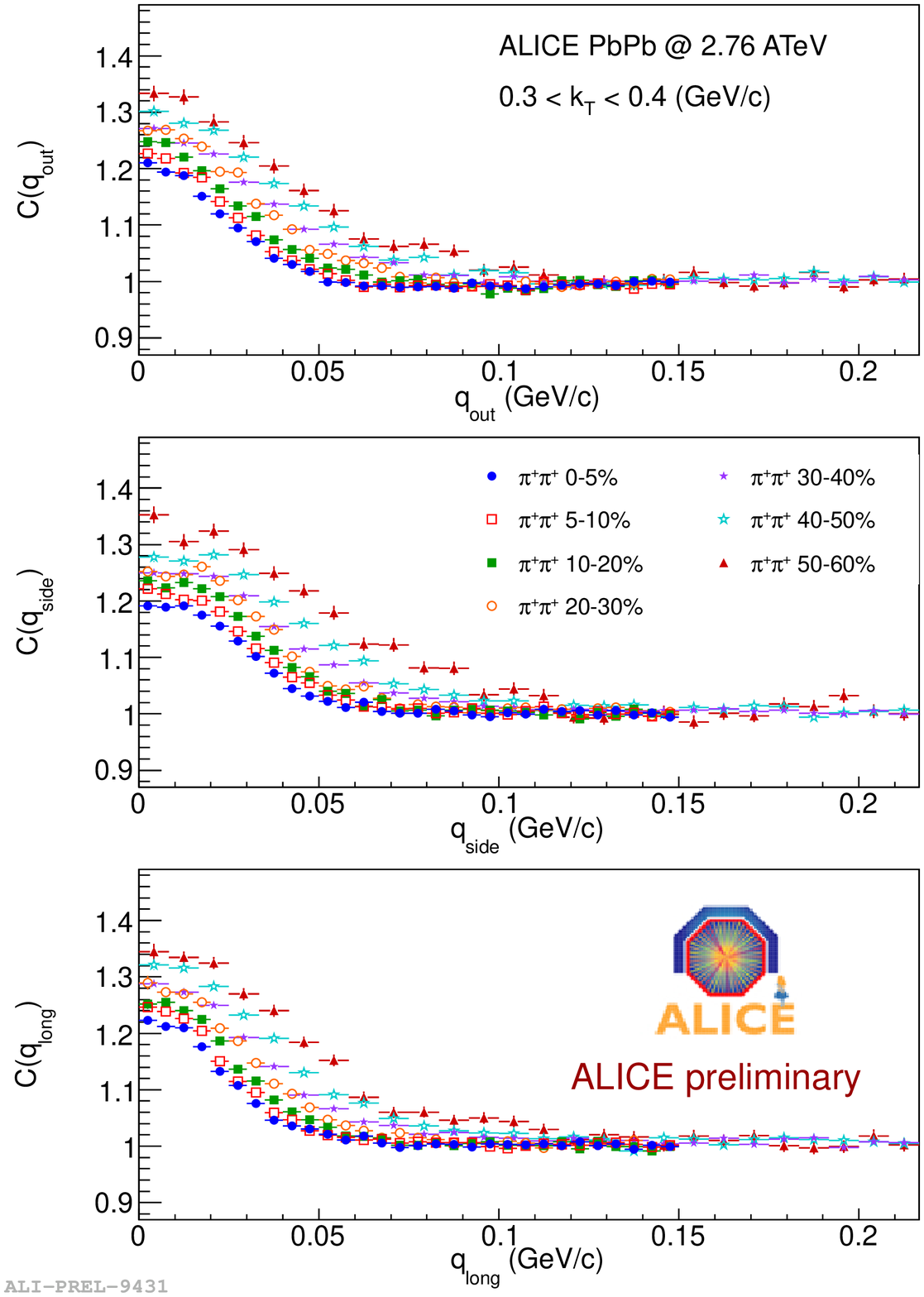}
\caption{Comparison of the Pb--Pb correlation functions in Cartesian representation for $\pi^+\pi^+$ for seven centrality classes.}
\label{fig:CFPBPBcent3D}
\end{minipage}
\hspace{0.5cm}
\begin{minipage}{.45\textwidth}
\includegraphics*[width=\textwidth]{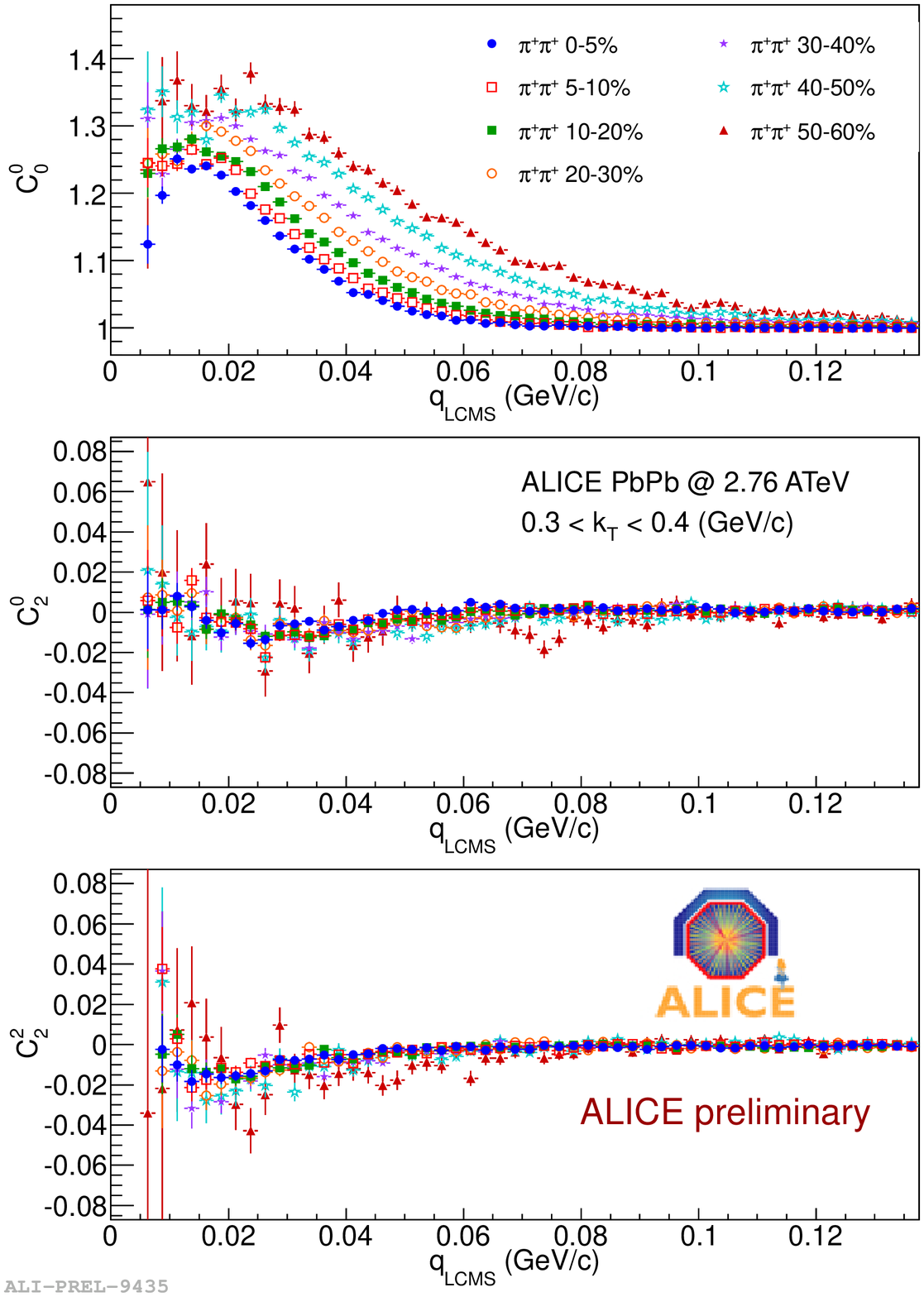}
\caption{Comparison of the Pb--Pb correlation functions in Spherical Harmonics representation for $\pi^+\pi^+$ for seven centrality classes.}
\label{fig:CFPBPBcentSH}
\end{minipage}
\end{figure}

The correlation functions for seven centrality classes and selected $k_{\mathrm{T}}$ range (0.3-0.4)~GeV/$c$ are shown for 3D Cartesian representation in Fig.~\ref{fig:CFPBPBcent3D} and for Spherical Harmonics representation in Fig.~\ref{fig:CFPBPBcentSH}.

\section{Results}
\label{sec:results}
The extraction of the femtoscopic radii was performed by fitting the Bowler-Sinyukov formula to the experimental data:
\begin{equation}
C_{\mathrm{f}}(\mathbf{q}) = (1-\lambda) + \lambda K_{\mathrm{C}}(q_{\mathrm{inv}}) \left [1+\exp(-R_{\mathrm{out}}^{2}q_{\mathrm{out}}^{2}-R_{\mathrm{side}}^{2}q_{\mathrm{side}}^{2}-R_{\mathrm{long}}^{2}q_{\mathrm{long}}^{2}) \right ],
\label{eq:cfun}
\end{equation}
where $long$ is the direction along the beam, $out$ along the pair transverse momentum, and $side$ perpendicular to the other two, while $q_{\mathrm{inv}}$ is the magnitude of the invariant pair relative momentum. The factor $K_{\mathrm{C}}$ accounts for two charged pion Coulomb repulsion. The parameters $R$ are femtoscopic radii (source sizes) in each of the three directions. Let us focus on the $C_0^0$ component. The femtoscopic effect, coming from the symmetrization of the two-pion wave function, is visible as the increase of the correlation function below $q=0.5\ \rm{GeV/c}$. The fall visible for the lowest relative momenta is due to the Coulomb repulsion, but it has little impact on the Bose-Einstein peak and, therefore, on the extracted HBT radii. 

\begin{figure}[!ht]
\centering
\begin{minipage}{.45\textwidth}
\includegraphics*[height=12cm,width=\textwidth]{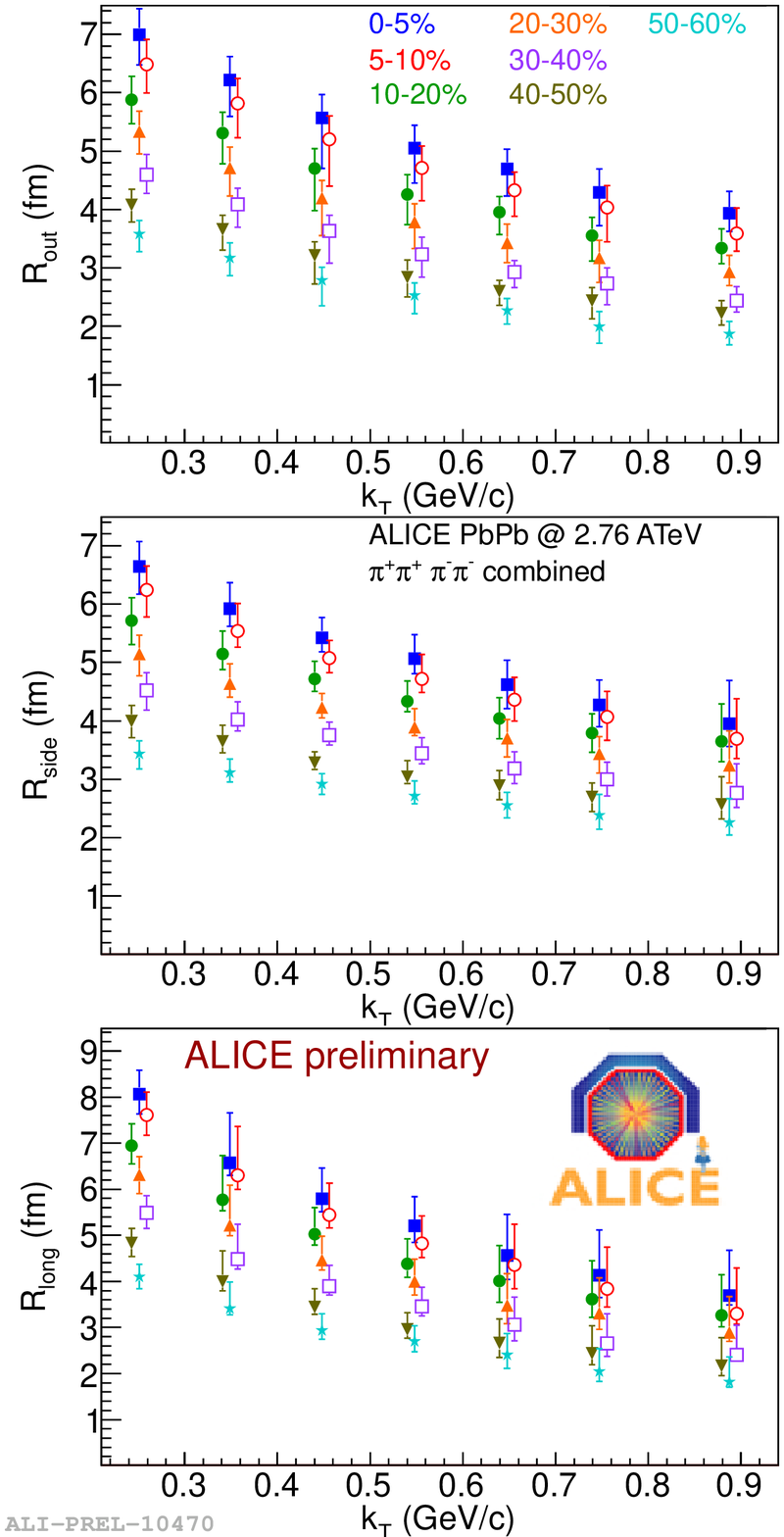}
\caption{Femtoscopic radii for Pb--Pb collisions as a function of pair momentum, for seven centrality ranges. The top, middle, and bottom panels show $R_{\mathrm{out}}$, $R_{\mathrm{side}}$, and $R_{\mathrm{long}}$ respectively. The error bars correspond to the combined statistical+systematic uncertainty.}
\label{fig:Radii}
\end{minipage}
\hspace{0.5cm}
\begin{minipage}{.45\textwidth}
\includegraphics*[height=12cm,width=\textwidth]{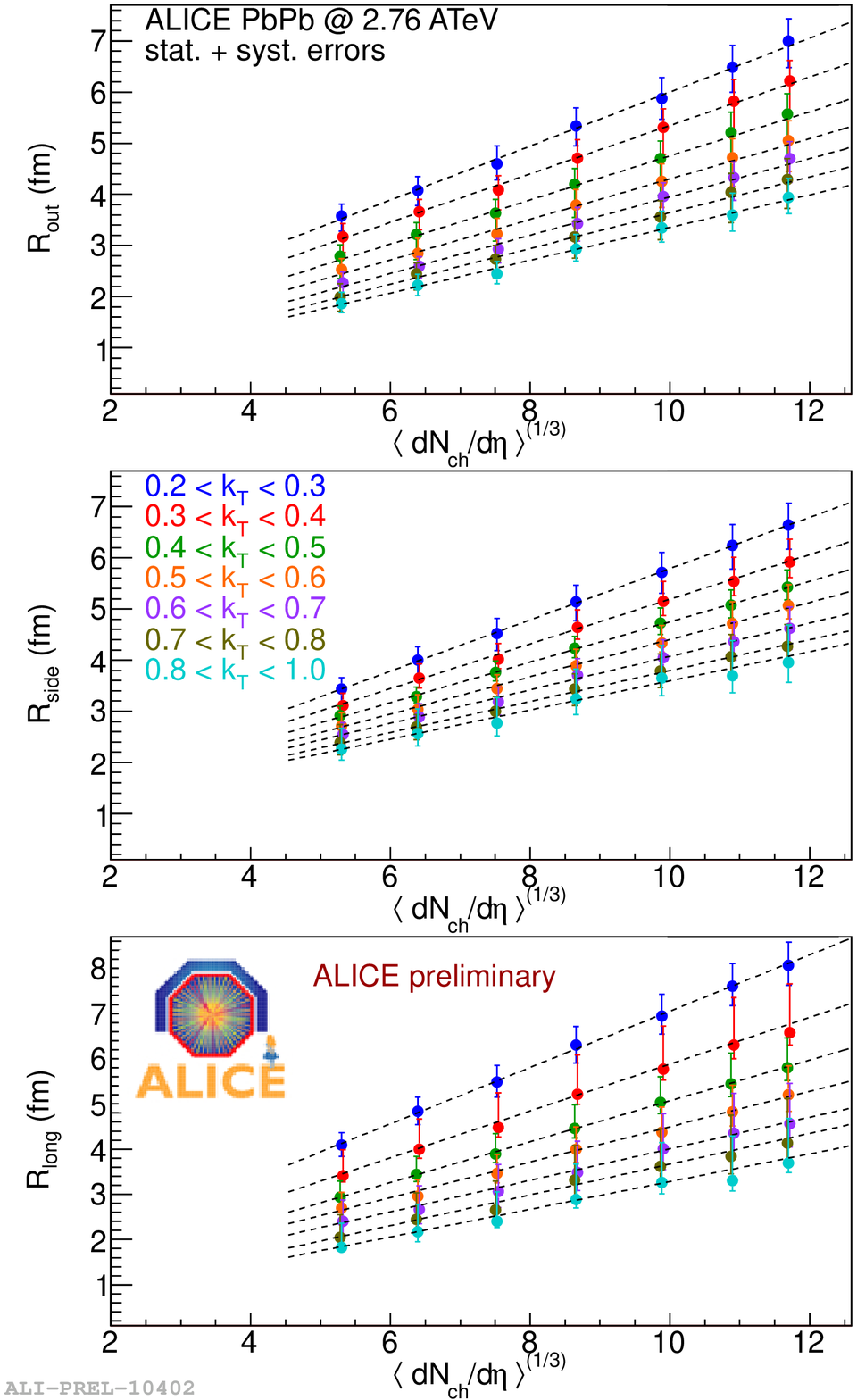}
\caption{Femtoscopic radii for Pb--Pb collisions as a function of event multiplicity, for even ranges in pair momentum. The top, middle, and bottom panels show $R_{\mathrm{out}}$, $R_{\mathrm{side}}$, and $R_{\mathrm{long}}$ respectively. The error bars correspond to the combined statistical+systematic uncertainty.}
\label{fig:RadiiCent}
\end{minipage}
\end{figure}

\begin{figure}[!htb]
\centering
\includegraphics*[width=10cm]{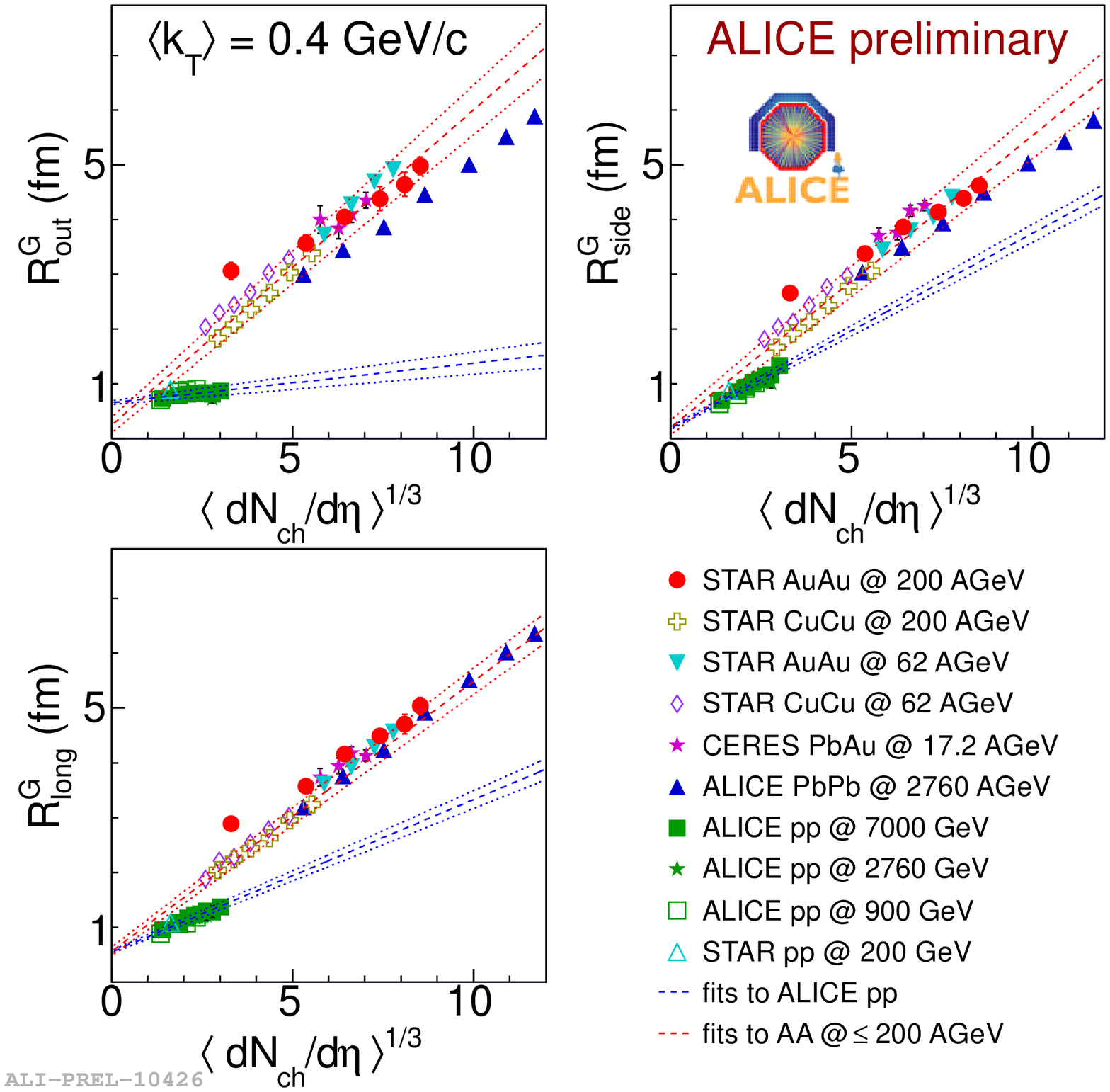} 
\caption[]{Comparison of femtoscopic radii, as a function of measured charged particle multiplicity, for many collision systems and collision energies. Lines show linear fits done separately to heavy-ion data and proton-proton data.}
\label{fig:pbtopcomp}
\end{figure}

The correlation functions for all centrality classes and pair momentum ranges have been fitted with Eq.~(\ref{eq:cfun}). The resulting femtoscopic radii as a function of $k_{\mathrm{T}}$ are shown in Fig.~\ref{fig:Radii}. All three radii, for all centrality ranges, show a power-law decrease with pair momentum. This dependence is qualitatively consistent with the mechanism arising in hydrodynamic modeling of heavy-ion collisions which is described in~\cite{Akkelin:1995gh}. 
Secondly, all three radii grow with increasing event multiplicity for all pair momentum ranges, which is better viewed in Fig.~\ref{fig:RadiiCent}. This is consistent with the expectation that the system with larger initial size, which later produces more final state particles, will also have larger size at freeze-out, to which the femtoscopic radii correspond. The exact nature of this scaling is illustrated in Fig.~\ref{fig:pbtopcomp}. The radii from Pb--Pb collisions scale linearly with cube root of the final-state charged-particle multiplicity, this scaling holds for all pair momentum ranges.



In Fig.~\ref{fig:pbtopcomp} the heavy-ion data from Pb--Pb collisions at the LHC are compared to various results obtained at lower collision energies, as well as to results from proton-proton collisions. It has been argued~\cite{Lisa:2005dd} that the 3-dimensional femtoscopic radii scale with cube root of measured charged particle multiplicity. The dashed lines in the figure are linear fits to heavy-ion (excluding the data from LHC) and \pp~data. The linear scaling is good for $long$ and $side$ directions and only approximate in $out$. Concerning the ALICE data the scaling in $long$ direction is preserved. 
The LHC data for the side direction go below the scaling trend determined by the low energy data, although they still agree considering the large statistical uncertainty. A clear departure from the linear scaling is seen in the $out$ direction; data from the LHC lie below the trend. Such behavior was predicted by hydrodynamic calculations~\cite{Kisiel:2008ws} and was the result of the modification of the freeze-out shape. Larger initial deposited energy produces larger temperature gradients and longer evolution time at LHC. This results in a change from outside-in to inside-out freeze-out and this modification of the space-time correlation drives the $R_{\mathrm{out}}/R_{\mathrm{side}}$ ratio to values lower than at RHIC. Therefore already for heavy-ion data in the transverse direction the simple linear scaling is broken.

\section{Conclusions}
We reported on the analysis of identical pion femtoscopic correlations in Pb--Pb recorded by ALICE. It was found that all three femtoscopic radii scale linearly as a function of cube root of charged particle pseudorapidity density and show the power-law dependence as a function of pair transverse momentum. The scaling was also compared to the ones observed in heavy-ion collisions at lower energies and to proton-proton collisions. It was found that the radii at LHC follow the scaling in the $long$ direction, but deviations are observed in the transverse directions. These were in qualitative agreement with predictions from hydrodynamic models. The scaling of the HBT radii is also seen for \pp~collisions at $\sqrt{s}=0.9$, 2.76, and 7~TeV. However, the scaling parameters are different from those for heavy-ion collisions and therefore, no simple scaling between \pp~and heavy-ion systems is present. The results suggest that and the initial state must be taken into account while interpreting the results.

\section*{Acknowledgements}
\label{sec:acknowledgements}
This work has been financed by the Polish National Science Centre under decisions no. DEC-2011/01/B/ST2/03483, DEC-2012/05/N/ST2/02757, and by the European Union in the framework of European Social Fund through the Warsaw University of Technology Development Programme, realized by Center for Advanced Studies.

\bibliography{bibliography}

\end{document}